\documentclass[12pt]{article}

\begin{document}

\newcommand{\ddx}[1]{\frac{\partial}{\partial x^{#1}}} 
\newcommand{\ddxp}[1]{\frac{\partial}{\partial x'^{#1}}} 
\newcommand{\dydx}[2]{\frac{\partial{#1}}{\partial{#2}}}
\newcommand{\dydxt}[2]{\frac{d{#1}}{d{#2}}}
\newcommand{\DD}{\Delta}
\newcommand{\GG}{\Gamma}
\newcommand{\tp}{\otimes}             
\newcommand{\ww}{\wedge}             
\newcommand{\la}{\langle}  
\newcommand{\ra}{\rangle}  

\newcommand{\vv}[1]{{\bf #1}}             
\newcommand{\tpp}{\tp\cdots\tp}       
\newcommand{\h}{{\mathcal H}}
\newcommand{\f}{{\mathcal F}}
\newcommand{\G}{{\mathcal G}}
\newcommand{\W}{{\mathcal W}}
\newcommand{\LL}{\Lambda}
\newcommand{\LLL}{\mathcal L}
\newcommand{\ppp}{\partial}
\newcommand{\til}[1]{\stackrel{\sim}{#1}} 

\newcommand{\be}{\begin{eqnarray}}
\newcommand{\ee}{\end{eqnarray}}
\newcommand{\bes}{\begin{eqnarray*}}
\newcommand{\ees}{\end{eqnarray*}}

\newtheorem{prop}{Proposition}[section]

\title{Covariant Extended Phase Space for Fields on Curved Background}
\author{Pankaj Sharan\\
Physics Department,\\ Jamia Millia Islamia, New Delhi 110 025, INDIA}

\maketitle
\abstract{It is shown that the nature of physical time requires the extended phase space
in mechanics to have a bundle structure
with time as the 1-dimensional base manifold and the phase space as the fiber.
Phase trajectories are sections whose tangent vectors
annihilate the 2-form $-d\Theta$ where $\Theta$ is the 
Poincare-Cartan 1-form $\Theta=pdq-Hdt$.
This bundle picture of the extended phase space is then applied to fields in 
a covariant, `directly'  Hamiltonian formalism without requiring a Lagrangian
as a starting point. In this formalism
the base manifold is the four-dimensional space-time with a Riemannian metric. 
The canonical momenta are differential 
1-forms for each field degree of freedom.
The Poincare-Cartan 4-form has the general structure $\Theta=(*p)\wedge d\phi-H$
where * is the Hodge star operator of the Riemannian metric.  
Allowed field configurations are sections of the bundle such that 
the 4-dimensional tangent spaces
to these sections, annihilate the 5-form $-d\Theta$.
Noether currents are calculated for symmetry fields and 
a new bracket analogous to the Peierls bracket is defined on the extended phase space
in place of the Poisson bracket. }

\vskip 5mm

\section{Introduction}   

Formulating a dynamical problem in classical mechanics is always `directly 
Hamiltonian'. Given a configuration space $Q$, its cotangent bundle
$T^*(Q)$ possesses a unique fundamental 1-form $p_idq^i$ (summation over $i$ implied). 
All one needs is a  Hamiltonian function $H(q,p)$ and the dynamics is fixed by 
the Poincare-Cartan (PC) form $p_idq^i-Hdt$. 

But a relativistic {\em field} theory, it is thought,
must always begin with a Lagrangian because relativistic invariance is 
spoiled by the momenta which are related to time derivative of fields.  A great deal of effort is required to re-establish relativistic invariance \cite{dirac}. 
The subject of this paper is a `directly Hamiltonian',  relativistically covariant 
field theory which offers additional insights.

The first step in this direction is to treat all the  four derivatives of field
 $\ppp_\mu \phi$ on the same footing. We can do that by writing $d\phi =\ppp_\mu\phi dx^\mu$
This is analogous to $dq$ of mechanics. Since for the simplest cases $p$ is proportional to
$\dot{q}$ in mechanics, we guess that we should {\em define the canonical momenta for scalar fields
to be differential 1-forms} like $d\phi$. 

The all-important quantity {\em action} is simply the integral of the
PC-form $\int(p_idq^i-Hdt)$. In field theory it is an integral over a four-dimensional domain.
Thus the PC-form for fields should be a differential 4-form. To write $p\ww d\phi$ will give us only
a 2-form. But if we use the metric of space-time, we can convert a 1-form $p$ into a 3-form
via the Hodge star operator. So we suggest the PC-form to be of the form $(*p)\ww d\phi-H$
where  $H$, the 'Hamiltonian'  is a suitable 4-form made solely out of $\phi$ and $p$ (or $*p$).
No $d\phi$ or $dp$ should be used in $H$. The 4-form $H$ should not to be confused with the energy 
density which is a 3-form having constant integral on space-like surfaces in static space-times.

The purpose of this paper is to outline a new formulation of fields in a
curved background. The main features of the formalism are as follows :
\begin{enumerate}
\item It is a totally covariant, directly Hamiltonian formalism in the extended phase space
(called EPS hereafter) which includes space-time as well as the fields and momenta.
There is no Lagrangian defined. Instead a covariant differential 4-form $H$ determines
the dynamics. 
\item EPS has the structure of a bundle with 4-dimensional space-time
as the base. The typical fiber is the phase space of field variables and canonical  momenta.
In this we are prompted by classical mechanics where the bundle structure
is forced by unidirectional nature of physical time \cite{bundle}. 
\item The field variables are vector space valued 
differential forms. The canonical momenta are differential forms of a degree
higher. 
\item The central objects of the theory are the Poincare-Cartan (PC) differential
4-form $\Theta$ (analogous to the 1-form $pdq-Hdt$ of
mechanics) and its derivative 5-form $\Xi=-d\Theta$. (The negative sign is just a convention.)
\item The quantity $\int \Theta$  over a 4-dimensional surface in the 
EPS is the action. we limit ourselves to 
4-dimensional surfaces
which are {\em sections}, that is, mappings from the space-time base into the bundle respecting the
fiber structure. 
\item Stationarity of action implies that the allowed field configurations are those sections or
surfaces whose tangent spaces annihilate the 5-form $\Xi=-d\Theta$. This condition is 
precisely equivalent to the field equations. 
We call these sections as `solution surfaces'.
This is completely analogous to 
Hamiltonian mechanics where phase trajectories annihilate the 2-form $-d(p_idq^i-Hdt)$.
\item Observables of the theory are generally of the form $\int A$ where $A$ is a 
4-form defined on the EPS. By suitably `smearing'  the 4-form $A$ with functions of compact support we can also include the local observables of field theory.
\item Fields and canonical momenta are forms of different degrees. 
There cannot be  a Poisson bracket in the usual sense. 
Instead, a covariant bracket defined by Peierls long ago
is the natural bracket in our formalism. 
Peierls bracket is a generalization of the Poisson bracket in the sense that 
equal-time Peierls brackets coincide with Poisson brackets when time is singled out
and time derivative of the field related to canonical momentum as in the usual `3+1'
formulation of field theory. 
\item The Riemann metric of the base manifold plays a crucial role in the formalism
through the Hodge star operator which converts $r$-forms into $(4-r)$-forms.
\item The formalism limits the kind of fields that can exist so far as their differential-geometric 
nature in space-time is concerned. Because of the star duality, the fields
can only be 0, 1 or 2-forms. However, there is no restriction on the internal space.
The theory can cover all known matter and gauge fields in arbitrary background space-time.   
\item Symmetries, including gauge symmetries are mappings on the EPS that leave the PC form invariant.
For infinitesimal symmetry mappings generated by a vector field $Y$ on EPS, the 3-form $i(Y)\Theta$
integrated on a closed 3-surface of the 4-dimensional solution surface is zero. This is simply
the statement of Noether's theorem.
\end{enumerate}

In the next section we review the extended phase space in mechanics. The requirement that
 physical time always moves forward reveals a bundle structure of the EPS.
The role of PC-form and its derivative 2-form $\Xi=-d\Theta$ is examined, particularly the
factorization properties of $\Xi$. In section 3 we begin to construct
the extended phase space and PC-form for fields for scalar fields. In section 4 Noether currents
are defined and calculated. Section 5 explains the concept of observables and the 
Peierls brackets in this formalism. The last section is devoted to briefly enumerating ideas
on variational principle for fields which have inspired the present formalism.

\section{Bundle structure in Classical Mechanics}

\subsection{Extended Phase Space}
There are two ways to look at the evolution of a classical mechanical system in 
the phase space. 

One, a phase space, at least locally, has a fundamental
form $\theta=p_idq^i$ defined on it. Its exterior derivative (with a conventional negative sign),
$-d\theta=dq^i\ww dp_i$, is a non-degenerate 2-form. In other words, its components are
an antisymmetric matrix with determinant non-zero. This 2-form
can be used to convert vector fields on the phase space into differential 1-forms 
by contraction.  The inverse to this  mapping associates a vector field
to a 1-form. For any function $f(q,p)$ on the phase space
the vector field associated with the 1-form $df$  is called the Hamiltonian 
vector field of $f$, and it is given by
\bes X_f=\dydx{f}{p_i}\dydx{}{q^i}-\dydx{f}{q^i}\dydx{}{p_i}.\ees
Integral curves of the vector field $X_H$ of
the Hamiltonian function $H(q,p)$ are the evolution trajectories.

Equivalently, one can formulate dynamics in the {\em extended} phase space \cite{arnold}. 
Here, time is taken as an additional dimension adjoined to the phase space. 
The fundamental form $\theta$ is 
augmented to the Poincare-Cartan 1-form (PC form) 
$\Theta\equiv p_idq^i-Hdt$ where $H$ is the Hamiltonian function.
A curve $C$ in this $(2n+1)$-dimensional space is called a {\em characteristic curve} 
of the 2-form $\Xi\equiv -d\Theta$ if the tangent vectors $X$ to the curve $C$
``annihilate $\Xi$'' : that is, as a 1-form
\bes i(X)\Xi=0,\ees 
where $i(X)$ denotes the interior product of the vector field $X$ with what follows after it.
The form $\Xi$ will play a central role in our formulation
The class of characteristic curves (or 1-dimensional sub-manifolds) provides solutions to the dynamical problem. 
Let $C: s\to (t(s),q(s),p(s))$ be a curve and $H(q,p)$ the Hamiltonian function.
The tangent vector to the curve is
\bes X= \dydxt{}{s}=\dydx{t}{s}\dydx{}{t}+\dydxt{q^i}{s}\dydx{}{q^i}+\dydxt{p_i}{s}\dydx{}{p_i}, \ees
and the derivative of the PC form is
\be \Xi=-d\Theta&=&dq^i\ww dp_i+dH\ww dt \nonumber\\
&=& dq^i\ww dp_i+\dydx{H}{q^i}dq^i\ww dt+\dydx{H}{p_i}dp_i\ww dt.
\label{-dtheta}
\ee
The interior product of $X$ with 2-form $\Xi$ is the 1-form
\bes i(X)\Xi &=&-\dydx{H}{q^i}\dydxt{t}{s}dq^i-\dydx{H}{p_i}\dydxt{t}{s}dp_i
+\dydxt{q^i}{s}dp_i\\
&+&\dydx{H}{q^i}\dydxt{q^i}{s}dt-\dydxt{p_i}{s}dq^i+\dydx{H}{p_i}\dydxt{p_i}{s}dt\\
&=&\left(\dydx{H}{q^i}\dydxt{q^i}{s}+\dydx{H}{p_i}\dydxt{p_i}{s}\right)dt
-\left(\dydx{H}{q^i}\dydxt{t}{s}+\dydxt{p_i}{s}\right)dq^i
\\&+&\left(-\dydx{H}{p_i}\dydxt{t}{s}+\dydxt{q^i}{s}\right)dp_i.
\ees
Thus $X$ annihilates $\Xi$ if
\be \dydxt{H}{s}&=&0,\\
\dydxt{q^i}{s}&=&\dydx{H}{p_i}\dydxt{t}{s},\\
\dydxt{p_i}{s}&=&-\dydx{H}{q^i}\dydxt{t}{s}.
\ee
These are equivalent to the canonical equations of motion
\bes \dydxt{q^i}{t}=\dydx{H}{p_i},\qquad \dydxt{p_i}{t}=-\dydx{H}{q^i}\ees
{\em provided} $(dt/ds)\neq 0$. 

The seemingly obvious requirement $dt/ds\neq 0$ is quite deep in fact.
Physically, $(dt/ds)>0$ (for example)
means that {\em time never stops and always moves forward} with parameter $s$. 
Mathematically,
$(dt/ds)>0$ suggests a bundle picture as follows. 

$dt/ds$ is the push-forward, under the projection $(t,q,p)\to t $, of the 
tangent vector to the 1-dimensional sub-manifold
$s\to (t(s),q(s),p(s))$ into the tangent vector to the projected curve $s\to t(s)$ . 

Therefore if the
physical time $t$ is chosen as the 
one-dimensional {\em base} manifold with the phase space with coordinates $(q,p)$
as fiber, and $\pi :(t,q,p)\to t$ the projection
then $C : s\to (t(s),q(s),p(s))$ is a ``projectable'' sub-manifold,
that is, a manifold whose (non-zero) tangent vectors map to non-zero vectors 
on the base by the push-forward map $\pi_*$. We can equivalently say that
the allowed trajectories in the extended phase space are ``sections'' $t\to (t,q(t),p(t))$
from the one-dimensional base space into the bundle which are, moreover, characteristics
of $\Xi $. In other words, the tangent spaces of the sub-manifold annihilate $\Xi$. 

It is important
to realize that the ``arrow of time'' (either $dt/ds>0$ for all $s$ or $dt/ds<0$ for all $s$)
is a consequence of the continuous nature
of the section. We shall see that causality, or the unidirectional nature of time for cause and effect,
remains an important ingredient for fields in extended phase space formalism as well.

\subsection{Variational Principle}
The characteristic curves of the 
two-form $\Xi=-d\Theta$ are related to the variational principle in 
the extended phase-space. Let $C: s\to (t(s),q(s),p(s))$ be as above
with $s\in(s_1,s_2)$. The {\em action}, defined
as the integral of the PC form on this curve is the quantity
\bes A(C)=\int_C\Theta=\int_C(p_idq^i-Hdt).\ees
Let $Y$ be a vector field in the extended phase space representing 
infinitesimal variations. $Y$ need only be defined in the neighborhood of the
(image of the) curve $C$. The action $A(C)$ is stationary if its Lie derivative 
$\LLL_YA(C)=\int \LLL_Y\Theta$ 
with respect to $Y$ is zero. Using the formula $\LLL_Y=i(Y)\circ d+d\circ i(Y)$
which is true when acting on differential forms, we calculate
\bes \LLL_Y(A(C))=-\int i(Y)\Xi+[\left.i(Y)(p_idq^i-Hdt)]\right|^{s_2}_{s_1}, \ees
where the second term, which comes from the exact differential $d \circ i(Y)\Theta$
is zero if variation of $q^i$'s and $t$ is zero on the boundary points at $s=s_1,s_2$. 
There is no restriction on variation of $p_i$'s however. The first term when evaluated
along $C$ will be the integral  of 
$ -(i(Y)\Xi)(X)$
where $X$ is the tangent vector along $C$.
Thus, 
\begin{quote}{\em 
for  variation field $Y$, which is arbitrary except for the boundary conditions 
noted above, action is stationary if and only if  $i(Y)\Xi$ evaluated on
the tangent to the curve $C$ is zero at all points on $C$. }\end{quote}

\subsection{Factorization Property}
In order to generalize to field theory we demonstrate a useful factorization property
of $\Xi$. Refer to equation \ref{-dtheta} above and write 
\be \Xi = \left(dq^i-\dydx{H}{p_i}dt\right)\ww\left(dp_i+\dydx{H}{q^i}dt\right) \ee
where a superfluous term proportional to $dt\ww dt=0$ has been introduced to obtain
the factorization.
The variational principle says that for arbitray variational field 
$Y$, the 1-form $i(Y)\Xi$ must vanish
on the proposed phase trajectory. If we choose $Y=\ppp/\ppp q^1$, for example, then 
\bes i(Y)\Xi = \left(dp_1+\dydx{H}{q^1}dt\right).\ees
Evaluated on the tangent vector to the trajectory $t\to (t, q^i=F^i(t),p_i=G_i(t))$ it gives
\bes \dydxt{G_1}{t}=-\left.\dydx{H}{q^1}\right|_{q=F,p=G}. \ees  
By choosing $Y$ in different directions of the phase space 
we get all the Hamilton's equations.

\section{Scalar fields}
Usually a field system is said to involve infinitely many degrees of
freedom. According to this traditional view each value $\phi(\vv{x},t)$ for space points $\vv{x}$
on a plane of constant time $t$ is a separate degree of freedom for a scalar field. This is 
the usual `3+1' Hamiltonian point of view. 
See Chernoff and Marsden\cite{chernoff} for a rigorous account of Hamiltonian
systems of infinitely many degrees of freedom. The global-geometric view taken by 
our approach is to regard
the entire field configuration over all space-time as a 4-dimensional surface `above'  space-time
in the extended phase-space. The infinitely many degrees of freedom appear
only when time is chosen as a special parameter of evolution. 

We now generalize the idea to the extended phase space
for  fields. The bundle picture remains, except that the 1-dimensional base space of time
is now replaced by the 4-dimensional space-time.
We consider the scalar field in arbitrary curved back ground to illustrate
the idea.

In field theory, the field $\phi$ is the configuration 
variable analogous to $q$. Time and space are four ``time'' variables $t^\mu,\mu=0,1,2,3$. 
The PC-form for fields is a differential
{\em four}-form whose space-time integral is the action. 
As discussed in the Introduction, for a scalar field $\phi$  
its canonical momentum is a differential  1-form $p=p_\mu dx^\mu$
and the PC 4-form looks like :
\be \Theta=(*p)\ww d\phi-H \ee
where the  `Hamiltonian' $H$ is a differential 4-form made from $\phi$ and $p$ only. 
We choose it to be
\be H &=& \frac{1}{2}(*p)\ww p+\frac{1}{2}m^2\phi^2(*1)\nonumber\\
&=& \left(-\frac{1}{2}\la p,p\ra+\frac{1}{2}m^2\phi^2\right)(*1).
\ee
We have used the definition of the star operator relating it to
the inner product determined by $g_{\mu\nu}$ because
\be *(dt^\mu)\ww(dt^\nu)=-g^{\mu\nu}(*1). \ee
Our notation is the same as Sharan\cite{sharan} or
Choquet-Bruhat and DeWitt-Morette\cite{AMP}.

For our case $\Xi$  is a differential 5-form and it can be calculated
easily. Using
\bes d(*p\ww p)=d(*p)\ww p+*p\ww(dp)=2d(*p)\ww p \ees
we get
\bes dH=(d*p)\ww p+m^2\phi\, d\phi\ww(*1). \ees
Substituting in $\Xi=-d\Theta $ we see that it factorizes :
\be \Xi = (d*p-m^2\phi(*1))\ww(p-d\phi), \ee
where we use the fact that the 5-form $(*1)\ww p$ 
in four variables $t$ is zero because there are five factors of $dt$'s.
This is completely analogous to the factorization of $\Xi$ in mechanics.

We now use our variational principle. Take $Y=\ppp/\ppp \phi$, and the 
interior product $i(Y)\Xi$ is the 4-form
\bes i\left(\dydx{}{\phi}\right)\Xi= -(d*p-m^2\phi(*1)).\ees
When evaluated on the section $t\to \phi(t),p_\mu(t)dt^\mu$ it gives
\bes -d*[p_\mu(t) dt^\mu] +m^2\phi(*1)=0.\ees
If $Y$ is chosen to be $\ppp/\ppp p_\nu$ then because
$ d*p=d*(p_\mu dt^\mu)=dp_\mu\ww *(dt^\mu)$ we have
\bes i\left(\dydx{}{p_\nu}\right)\Xi &=& *(dt^\nu)\ww \left(p_\mu(t) - \dydx{\phi}{t^\mu}\right)dt^\mu\\
 &=& -g^{\nu\mu}\left(p_\mu(t) - \dydx{\phi}{t^\mu}\right)\\ &=& 0. \ees
This gives us four equations
\be p_\mu- \dydx{\phi}{t^\mu}=0. \ee
Thus (now treating $\phi$ and $p$ as functions of $t$) we get the 
Hamiltonian equations
\be p=d\phi,\qquad d*p-m^2\phi (*1)=0. \ee
The second of these equations is the Klein-Gordon equation for $\phi$ 
on curved background when the first is substituted in it because
\be d*d\phi &=& \ppp_\mu(\sqrt{|g|}g^{\mu\nu}\ppp_\nu\phi)
dt^0\ww\dots\ww dt^3\nonumber\\
&=& \frac{1}{\sqrt{|g|}}\ppp_\mu(\sqrt{|g|}g^{\mu\nu}\ppp_\nu\phi)(*1)
\ee

\section{Noether Currents}
Let $D$ be a domain in the 4-dimensional base of space-time.
Let $\sigma $ be the mapping from the base into the bundle.
Then action for this section is 
\be A(\sigma)=\int_{\sigma(D)}\Theta.\ee	
The variational principle for fields (in complete analogy to mechanics) can be 
written as
\be \delta_Y A(\sigma) 
&=&  \int_{\sigma(D)}\LLL_Y \Theta \nonumber \\ 
 &=&  \int_{\sigma(D)}d[i_Y\Theta]  -\int_{\sigma(D)}i_Y\Xi \nonumber\\
 &=& \oint_{\ppp \sigma(D)}i_Y\Theta -\int_{\sigma(D)}i_Y\Xi\ee
where $Y$ is the variational field in the EPS. If $\phi$ is kept fixed on the boundary then
$Y$ has zero component in the direction of $\phi$ and $i(Y)\Xi=0$. Consequently,
the first term is zero and the equations of motion are obtained 
from $\delta_YA(\sigma)=0$
as the condition
that the 4-form $i(Y)\Xi$ should vanish when evaluated on the tangent vectors 
to the 4-dimensional section or solution-surface. This is the criterion we have used.

On the other hand if we already have  a solution surface 
and $Y$ is arbitrary  then the second term is zero and the variation is
\be \int_{\sigma(D)}\LLL_Y \Theta =\oint_{\ppp \sigma(D)}i_Y\Theta .
\ee

A symmetry transformation is given by a field $Y$ such that 
$\LLL_Y \Theta =0$. Then for such symmetry transformations
\be \oint_{\ppp \sigma(D)}i_Y\Theta =0. \ee
The 3-form $i_Y\Theta $ is called the Noether current and the statement above is 
the conservation law. If the boundary $\ppp(\sigma(D))$ is taken to be 
of the shape of two space-like surfaces joined together at spatial infinity.
Then the surface integral involves the $dt^1\ww dt^2\ww dt^3$
component of the Noether 3-form. This is the more usual statement of the Noether theorem. 

Usually, the symmetry fields satisfy the stronger conditions $L_Y(*p\ww d\phi)=0$ and 
$L_Y H=0$ separately.

As an example, if 
\bes Y=v^\mu(t)\dydx{}{t^\mu}\ees
then the Noether current 3-form is given by,
\be
i_Y\Theta &=&
\left[\phi_{,\mu}\phi_{,\nu} -\frac{1}{2}g_{\mu\nu}\left(g^{\alpha\beta}\phi_{,\alpha}\phi_{,\beta}
+m^2\phi^2\right)\right]v^\mu*(dt^\nu)\nonumber \\
&&.
\ee
While calculating these currents one must not assume the fields $\phi$ and $p$ 
to be on the solution surface to begin with. Only after taking interior product (that is contracting)
with $Y$ can the field be evaluated at the solution surface. The details of this not entirely
trivial calculation are given in Appendix A. For Minkowski space-time admitting
translational symmetry the $dt^1\ww dt^2\ww dt^3$ components of conserved 
quantities (energy and momentum densities) are the familiar expressions given below.
\vskip 5mm
\begin{center}
\begin{tabular}{ccl} $Y$ & $v^\mu$ & Coefficient of $-dt^1\ww dt^2\ww dt^3$ in $i_Y\Theta$\\ 
&&\\
$\ppp/\ppp t^0$ & $(1,0,0,0)$ & $(1/2)[(\phi_{,0})^2+(\nabla\phi)^2+m^2\phi^2]$\\
&&\\
$\ppp/\ppp t^1$ & $(0,1,0,0)$ & $\phi_{,1}\phi_{,0}$\\
\end{tabular}
\end{center}

\section{Observables and Peierls bracket}

Our formalism treats coordinate $\phi$ and its canonical momentum $p$ respectively
as 0- and 1-forms. In classical mechanics they seem to be quantities of the 
same type because in one-dimensional base manifold representing time, 
0-forms and 1-forms are both 1-dimensional spaces. 
This situation changes for field theory in four dimensions. There 0- and 1-forms 
are respectively spaces  of one and four dimensions.

We define observables of our theory to be 
integrated quantities over a four dimensional sub-manifold of the EPS.
Quantities like action are a good example. 
A typical observable is determined by a 4-form $A=\int \alpha$.
The support of $\alpha$, that is set over which it has non-zero values could be  
suitably restricted to allow for local quantities as observables. For
example, the scalar field $\phi$ is related to the observable
$\int \phi j (*1)$   where $j(t)$ is a scalar `switching function' 
which is non-zero in a small space-time region. 
For simplicity we would call both the
integrated as well as the non-integrated quantity by the same name `observable',
and it leads to no confusion.

The Peierls bracket \cite{peierls} , promoted extensively by DeWitt \cite{dewitt}, is the natural bracket-like quantity in 
this formalism. When the Hamiltonian 4-form $H$ is perturbed by observable $\lambda B$
(where $\lambda$ is an infinitesimal parameter)
the solution manifold shifts, and, after taking causality into account,
the difference between the two solutions at different points in the limit of $\lambda\to 0$ 
determines a `vertical' vector field $X_B$. This field
changes all other observables. The change in an observable
$A$ is equal to the Lie derivative $D_BA\equiv L_{X_B}A$ of $A$ with respect to $X_B$. 
Switching the roles of $B$ and $A$ we can calculate $D_AB$. The
Peierls bracket $[A,B]$ is defined as the difference $D_BA-D_AB$. 

For illustration we outline the calculate the Peierls bracket for the scalar field
with itself in Minkowski space.
The observable in question is the integrated 4-form 
\bes B=\int \beta=\int \phi j (*1) \ees 
where $j$ is a switching function in space-time with which the field $\phi$ is `smeared'. The 
Hamiltonian is changed to $H+\lambda B$ and the solution
manifold given by $t\to \phi=F_0(t),p_\nu=F_{0,\nu}$ gets
modified to a solution manifold which is determined by the 5-form
\bes \Omega_B &=& -d(*p)\ww d\phi +dH +\lambda d\phi j (*1)\\
&=& [d(*p)-m^2\phi(*1)-\lambda j(*1)]\ww [p-d\phi]. \ees
No derivative of $j$ appears because that would involve five factors of $dt$'s
and there can be only four such factors in a wedge product.
The equations for a solution $t\to \phi=F(t),p_\nu=G_\nu$ become
\bes G_\nu=F_{,\nu},\qquad (\ppp^\mu\ppp_\mu -m^2)F = \lambda j. \ees
The modification caused by $\lambda B$ as $\lambda \to 0$ to the solution 
$F_0$ is given by the retarded solution to the inhomogeneous Klein-Gordon equation,
\bes F(t)=F_0(t)+\lambda K(t),\qquad G_\nu=F_{,\nu} \ees 
where
\bes K(t)=\int G_R(t-s)j(s)d^4s. \ees
The retarded and advanced Green's functions $G_R(t),G_A(t)$ are the unique solutions
\bes G_{R,A}(t)=\frac{1}{(2\pi)^4}\int d^4k\, 
\frac{\exp(-ik^0t^0+i\vv{k}\cdot\vv{t})}{(k^0\pm i\epsilon)^2-\vv{k}^2-m^2}\ees
of
\bes (\ppp^\mu\ppp_\mu -m^2)G_R(t) =\delta^4(t)\ees
with the boundary condition that $G_R(t)$ is non-zero only in the forward light-cone
and $G_A(t)$ in the backward light-cone.

Thus the vertical field is determined to be ($\lambda\to 0$ can be factored out
to give the tangent vector field)
\bes Y_B=K(t)\dydx{}{\phi}+K_{,\nu}\dydx{}{p_\nu} \ees

Consider the observable 
\bes A =\int \alpha=\int \phi k (*1) \ees 
where $k(t)$ is another switching function. The change in $A$
due to $B$ is given by $D_BA =L_{Y_B}(A)$. Now,
\bes L_{Y_B}(A) &=& \int [i_Y (d\phi k (*1))+d(\phi k \,i(Y)(*1))]\\
&=& \int k K (*1), \ees
because $i(Y_B)(*1)=0$. Thus 
\bes D_BA &=&\int d^4t k(t)K(t)\\
&=& \int\int d^4t d^4s k(t)G_R(t-s)j(s) 
\ees
Reversing the role of $B$ and $A$ we get the Peierls bracket
\bes [A,B]=D_BA-D_AB=\int\int d^4t d^4s k(t)\Delta(t-s)j(s) \ees
where $\Delta$ is the Pauli-Jordan function $\Delta=G_R-G_A$.
This is equivalent to the commutator
\bes [\phi(t),\phi(s)]=\Delta(t-s) \ees
when $k$ and $j$ are Dirac deltas with support at $t$ and $s$ respectively. 

The Peierls bracket for the field $\phi$ and momentum $p$ can be 
calculated by considering the observable 
\bes C=\lambda (*p)\ww l=-\lambda p_\mu l^\mu (*1)\ees
where in this case we
must employ a 1-form switching function $l$ to smear the momentum. The
5-form is 
\bes \Omega_C= [d(*p)-m^2\phi(*1)]\ww[p+\lambda l-d\phi]. \ees
The relevant equation for the modified solution is 
\bes (\ppp^\mu\ppp_\mu -m^2)F = \lambda \ppp^\mu l_\mu \ees
because $d(*p)$ becomes $d(*(d\phi-l))=\ppp^\mu\ppp_\mu\phi-\ppp^\mu l_\mu$.
The change in $B$ is
\bes D_CB &=& \int\int d^4t d^4s j(t)G_R(t-s)(\ppp^\mu l_\mu)(s). \ees
On the other hand we have already calculated the vertical field for $B$
which gives
\bes D_BC &=&-\int K_{,\mu}l^\mu (*1)\\ 
&=& -\int\int d^4t d^4s\, l^\mu(t) \ppp_{t^\mu} G_R(t-s)j(s)\\
&=& \int\int d^4t d^4s (\ppp_\mu l^\mu)(t)  G_R(t-s)j(s)\\
&=& \int\int d^4t d^4s  j(t)G_A(t-s)(\ppp_\mu l^\mu)(s)\ees
after integrating by parts in the third step.
Therefore, 
\bes [B,C]=\int\int d^4t d^4s  j(t)\Delta(t-s)(\ppp_\mu l^\mu)(s)\ees
which, for $j(t)=\delta^4(t)$ and $l_\mu=(1,0,0,0)\delta^4(s)$ gives the 
equal-time ($t^0-s^0$) canonical Poisson bracket
of the ``3+1'' version of field theory
\bes [\phi(t,\vv{t}),p_0(t,\vv{s})]=\delta(\vv{t}-\vv{s})\ees
because
\bes \ppp_0\Delta(t)=-\delta^3(\vv{t}). \ees

\section{Discussion and conclusion}

The mathematical formalism of this paper is 
similar to the ``multi-symplectic'' Lagrangian approach to field theory in the works 
of Le Page, as reviewed and developed by Kastrup\cite{kastrup}, the De Donder-Weyl\cite{rund}
approach of Kanatchikov\cite{kanat} and the covariant
Hamiltonian-Jacobi formalism of Rovelli\cite{rovelli}. Recent contributions to multisymplectic formalism
are by Gotay and collaborators\cite{gotay}.
Our approach is different from these because
we use the background space-time metric
in an essential way through the Hodge star operator. Also, we treat the
space-time degrees of freedom $t^\mu$ (which specify the base) very differently 
from the field or momentum degrees
of freedom which are in the fibre above the base.
We require the PC-form to be a 4-form whose first term is linear in $d\phi$ to imitate $pdq$ term
and the second term is a 4-form $-H$ proportional to volume form $(*1)$. 
If, for instance, there are two fields $\phi_1$ and $\phi_2$, a 
4-form involving a factor $d\phi_1\ww d\phi_2$ is possible in principle but that
does not seem be allowed in the formalism for matter fields. Similarly 
other `non-canonical' expressions are possible in place of the standard $pdq-Hdt$ like
expression. For gravity, the Einstein-Hilbert
PC form does seem to have a non-standard expression as we shall see in a later paper.
But gravity is a special case anyway. 
For gravity the `internal' degrees of freedom in the fibre related to arbitrary choice of
local inertial frames and space-time bases which define the 
transformation of all field and momenta differential forms happen to coincide.  

It is natural and tempting to put our formalism in the topologically non-trivial
situations but we avoid
that in this first of a series of papers and limit to clarifying the physical concepts. 
Thus we assume the 
bundle to be a cartesian product of space-time and the fibre manifold.

Our aim has been to develop a purely Hamiltonian approach and define 
a suitable bracket to help build a quantum theory. The only reliable way to
convert a classical theory into a quantum theory is to define a suitable antisymmetric 
(or symmetric) bracket for observables of the theory which can be re-interpreted 
in quantum theory as a commutator
(or anti-commutator). Our phase space has a very different character
than the traditional phase space and our coordinate and momenta are differential
forms of different degrees. In the traditional formalism the 
observables are real valued functions on the phase space
and the definition of the Poisson bracket uses the pairing of one coordinate 
with one canonical momentum degree of freedom. But that is special to one-time
formalism of mechanics.

Whereas the Hamiltonian vector field for any observable
exists for in mechanics, the same may not be so for fields. We have seen that the
concept of a covariant bracket introduced by Peierls\cite{peierls} in 1952 is a natural  object to use in our Hamiltonian theory of fields. Here
the rate of change of one quantity is taken when {\em the other quantity is added to the 
Hamiltonian as an infinitesimal perturbation} and vice-versa. The Poisson brackets of mechanics can be 
defined without reference to any Hamiltonian whereas the Peierls bracket
requires the existence of a suitable governing Hamiltonian. Roughly speaking, the Poisson bracket 
can be described as the ``equal time'' Peierls bracket with zero Hamiltonian. 

This gives us added insight into the Hamiltonian mechanics of one time formalism,
particularly the concept of causality in systems with time dependent Hamiltonians \cite{sharan2}.
The interesting features for one-time formalism of classical mechanics 
relating to causality and Peierls bracket which  are revealed by our formalism of fields
will be published elsewhere.

\begin{appendix}
\section{$i_Y\Theta$ for $Y=v^\mu\ppp/\ppp t^\mu$}
As an illustration of the Noether theorem in our formalism
let us evaluate $i_Y\Theta$ for the present scalar field case 
for space-time translations. The vector field for constant infinitesimal
displacement $v^\mu$ is
\bes Y=v^\mu\dydx{}{t^\mu}\ees
We are not assuming that space-time is flat or that $Y$ are Killing fields of translation
symmetry.

We know that 
\bes *p &=& p_\mu *(dt^\mu)\\
&=& \frac{1}{3!}\sqrt{-g}p_\mu g^{\mu\alpha}\varepsilon_{\alpha\nu\sigma\tau}
(\nu\sigma\tau)\\
&\equiv & \frac{1}{3!}\sqrt{-g}p^\alpha\varepsilon_{\alpha\nu\sigma\tau}
(\nu\sigma\tau)
\ees
where we introduce a convenient notation
\bes (\nu\sigma\tau) \equiv dt^\nu\ww dt^\sigma\ww dt^\tau, \ees
with similar notation for two or four factors of $dt^\mu$ and we have defined the
contravariant canonical momentum 
\bes p^\mu=g^{\mu\nu}p_\nu .\ees

A simple calculation using 
\bes i_Y(dt^\mu\ww dt^\nu\ww dt^\sigma) &=& v^\mu(dt^\nu\ww dt^\sigma)
-v^\nu(dt^\mu\ww dt^\sigma)\\&&+
v^\sigma(dt^\mu\ww dt^\nu)\ees
gives,
\bes i_Y(*p)=\frac{1}{2!}\sqrt{-g}p^\alpha v^\beta
\varepsilon_{\alpha\beta\sigma\tau}(\sigma\tau).\ees
We can write this  also as 
\bes i_Y(*p)= p_\mu v_\nu *(dt^\mu\ww dt^\nu)= *(p\ww Y^\flat)\ees
where $v_\mu=g_{\mu\nu}v^\nu$ and $Y^\flat=v_\nu dt^\nu$ is the covariant field
corresponding to $Y$ after lowering the index by the metric.

As $Y$ involves $\ppp/\ppp t^\mu$ whose action on $d\phi$ is zero
\bes i_Y(*p\ww d\phi)=(i_Y*p)\ww d\phi, \ees
and 
\bes i_Y[*p\ww p\,]&=&[i_Y*p]\ww p-*p(i_Yp)\\
&=&*(p\ww Y^\flat)\ww p-p(Y)*p.
\ees

The formula $i_Y(*p)=*(p\ww Y^\flat)$,
although elegant, is not very useful for calculations. A straightforward expression 
for $i_Y(*p)\ww p$ is
\bes i_Y(*p)\ww p=[p_\mu(p.v)-v_\mu(p.p)]*(dt^\mu) \ees
where
\bes p.v=p_\mu v^\mu=\la p,Y^\flat\ra,\qquad p.p= p_\mu p^\mu=\la p,p\ra. \ees

Thus the calculation of $i_Y\Theta$ proceeds as follows,
\bes i_Y\Theta&=&i_Y\left[*p\ww d\phi-\frac{1}{2}*p\ww p
-\frac{1}{2}m^2\phi^2*(1)\right]\\
&=&(i_Y*p)\ww\left(d\phi-\frac{1}{2}p\right)+\frac{1}{2}p(Y)*p\\
&&-\frac{1}{2}m^2\phi^2i_Y*(1)
\ees
Evaluating it on the solution surface means we can put $p=d\phi$. Using expression for 
$i_Y(*p)\ww p$, $p(Y)=p.v$ and the fact that  
\bes i_Y*(1)&=&\sqrt{-g}(v^0[123]-v^1[023]+v^2[013]-v^3[012])\\
&=& v_{\mu}*(dt^\mu),
\ees
we get
\bes
i_Y\Theta &=&\left(\frac{1}{2}[p_\mu(p.v)-v_\mu(p.p)]+\frac{1}{2}(p.v)p_\mu\right) *(dt^\mu)\\ 
&&-\frac{1}{2}m^2\phi^2v_\mu*(dt^\mu)\\
&=&\left.\left(p_\mu(p.v)-\frac{1}{2}\left[(p.p)
+m^2\phi^2)v_\mu \right]\right) *(dt^\mu)\right|_{0}\\
&=&
\la d\phi,Y^\flat\ra (*d\phi) -\frac{1}{2}\left[\la d\phi,d\phi\ra +m^2\phi^2\right](*Y^\flat)
\ees
which can also be written in the useful form
\be
i_Y\Theta &=&
\left[\phi_{,\mu}\phi_{,\nu} -\frac{1}{2}g_{\mu\nu}\left(g^{\alpha\beta}\phi_{,\alpha}\phi_{,\beta}
+m^2\phi^2\right)\right]v^\mu*(dt^\nu)\nonumber \\
&&
\ee

\end{appendix}

\end{document}